\documentclass[intlimits,twoside,a4paper]{article}

\usepackage{amsmath,amssymb}
\usepackage{graphicx}
\usepackage{wrapfig}

\usepackage[T2A]{fontenc}
\usepackage[cp1251]{inputenc}

\usepackage[eqsecnum]{cmpj3}

\issue{2020}{23}{3}{33603}
\doinumber{10.5488/CMP.23.33603}
\title[Degenerate behavior under nonconfinement]%
{Self-assembly in rod/coil block copolymers: Degenerate behavior under nonconfinement %
}
\author[X.-G. Han, N. Liang, H. Zhang]{X.-G. Han,
 N. Liang, H. Zhang }
\address{School
of Science, Inner Mongolia
 University of Science and Technology, Baotou 014010, China }
\date{Received January  13, 2020, in final form May 8, 2020}

\begin{document}

\maketitle

\begin{abstract}
The self-assembly of block copolymers containing rigid blocks have received abiding attention due to its rich phase behavior and potential for use in a variety of applications. In this work, under asymmetric interactions between rod/coil components, the self-assembly of coil/coil/rod ABC triblock copolymers is studied using self-consistent field of lattice model. In addition to micelles, centrosymmetric lamellae (CSLM), lamellae, perforated lamellae, strips and gyroids, non-centrosymmetric (NCSLM) lamellae and wavy  morphologies are observed as stable phases.  The phase diagram of  interaction between rod and coil components versus the rod fraction is constructed given a fixed interaction between coil components. For intermediate rod fraction, degenerate behavior is observed.  NCSLM and CSLM are degenerate structures. It is found that the entropy of chain conformation plays an important role for this rich behavior. A mechanism of the degenerate behavior  is proposed in coil/rod block copolymers under noncofinement. This study provides some new insights into the degenerate behavior of block compolymers, which can offer a theoretical reference for related experiments.

\keywords rod/coil block copolyer, degenerate  behavior, self-consistent field
%
\end{abstract}

\section{Introduction}

The self-assembly of coil/rod block copolymers has attracted
special attention in recent years due to their rich mesoscopic
structures and their tunable physical properties \cite{Horsch2005,Horsch2010,Jiang2013,Jiang2013a}. Various
morphologies including lamellae, strips, cylinders,
vesicles, perforated lamellae, arrowhead, wavy lamellae, zigzag
lamellae, and so on have been observed, as summarized in
several recent reviews \cite{Olsen2008,Hoeben2005,Wang2011}.
The different self-assembled structures of block copolymers are
of interest for a wide
field of applications such as photonic crystals, polymeric solar cells,
high density storage media, nanoporous membranes, etc. \cite{Park1997,Li2000,Thurn-Albrecht2000,Kim2001,Cheng2001,Black2001,Fasolka2001,Hufnagel2015,Deng2003}.

For  mesoscopic structures
of block copolymers of technological
applications, it is desirable to develop different routes
of engineering new block copolymer structures. Confinement  from the
boundaries influences the self-assembly process and can
generate novel mesostructures. A multitude of novel morphologies from the self-assembly of
confined block copolymers have been observed, such as single helices and stacked toroids \cite{Yu2006,Yu2011,Li2006,Liu2017}. Moreover, it has been found
that degenerate structures were frequently observed under some confined
conditions \cite{Yu2006,Yu2008,Yu2011,Chen2006,Chen2007,Chen2008}. For instance, a set of degenerate structures were found by Yu et al. at a
certain pore size under cylindrical confinement \cite{Yu2008}. Later, it was
found that the high symmetry of a spherical pore can enable getting
more complicated multiple-degenerate structures. When
the copolymer was confined in an ellipsoidal pore, the observed
morphologies were intermediate between those observed in
a cylindrical and in a spherical pore \cite{Yu2011}. It is concluded that the degenerate behavior of assembled structures is dependent on the degree of confinement of the system.

Some properties and behaviors of coil/rod  block copolymers were
not observed in coil/coil systems, such as the emergences of wavy and zigzag lamellae. 
Compared with lamellar phase, lengthened rod/coil interface in zigzag
morphology is obviously unfavorable to the free energy, while allowing the coils to become less stretched and hence suffering
less entropy penalty. When changing coil/rod AB diblock into coil/rod/coil  ABA  triblock copolymers by adding a coil block, the range of the region for lamellar structure in the phase diagram decreased evidently \cite{Chen12007} compared with  coil/rod diblock copolymers \cite{Chen12006}. In linear coil/rod AB multiblock block copolymers, this variation of self-assembled structures from 1-D lamellar, 2-D rectangular to 2-D hexagonal lattices was observed, at an identical rod-to-coil volume ratio \cite{ Lee2001}. These   phase behaviors originate from
the liquid crystal behavior of the rods and from the complex entropic
interactions of the flexible coils. The entropy of these systems is maximized only by the particular way decreasing the  grafting density of the separating rod and coil segments to maximize the free volume of the coils \cite{Lee2000,Lee2001a}. It is due that the existence of rigid block for the copolymer to some extent restricts the free volume of the coil blocks and  is not favorable to the entropy \cite{Chen12008}. Recently, it is demonstrated that the entropy plays an important role in the  formation of degenerate structure \cite{Wang2016}.  Whether the inner confinement produced by rigid blocks in rod/coil block copolymers leads to the emergence of the degenerate behavior remains to be clarified so far.

Theories and simulations which are important tools of the insights for
understanding the phase behavior of complex block
copolymers have been extensively applied. For example, Horsch et al. studied the self-assembly of coil/rod diblock copolymers using Brownian dynamics method, \cite{Horsch2005,Horsch2006, Horsch2006a,Horsch2010} and predicted that the rods self-assemble into hexagonally arranged chiral cylinders \cite{Horsch2005,Horsch2010}. Meanwhile, a few groups have identified the stability of different morphologies using the self-consistent field theory (SCFT) \cite{Li2009,Chen2007} that is complementary to
these methods because of its capability to accurately determine the free energy of each structure under mean field treatment for many-body systems.
Combined with other theoretical methods containing Flory's lattice theory, \cite{Masten1998} Maier-Saupe theory
\cite{Pryamitsyn2004} and Onsager theory \cite{Song2009} to account for the aligning interactions between rods, SCFT can also be used to study coil/rod block
copolymer systems. Another  Li and Gersappe group \cite{Li2001} studied
the self-assembly of coil/rod diblock copolymers by proposing
a lattice-based SCFT simulation where the aligning interactions
between the rods can be neglected by employing the rotational
isomeric state scheme.  In this report, a SCFT lattice model is used
\cite{Chen12006} and the rigid section of the copolymer is described similar to Li
and Gersappe's method.

In our previous
works we  used the SCFT lattice model to study the phase
behaviors in associative polymer
solutions \cite{Han2010,Han2011,Han2012,Han2015} and in coil-rod-coil ABC  triblock copolymers \cite{Han2017}. The effect of asymmetry of the coil block on the microphase separation was focused on a series of  fixed rod fractions.  The results showed that, for different coil fractions, an increase in the interaction between different components leads to different transitions between cylinders and lamellae. In this work, under asymmetric interaction between coil/rod components, the self-assembly of coil/coil/rod ABC triblock copolymers was studied using a self-consistent field of lattice model. The phase diagram of interaction between rod and coil components versus the rod fraction was constructed. The degenerate behavior of NCSLM and CSLM  was observed. The mechanism of the degenerate behavior and the stability of NCSLM is analysed.

\section{Theory\label{sec2}}
We consider $n$ linear  coil/coil/rod  ABC triblock copolymers in a cubic
lattice, and the degree of polymerization of each chain is $N$.  Each A, B and C segment which have the same size  occupies one lattice site, thus the total number of the lattice sites $N_\text{L}$
equals $nN$.  The free energy functional of $F$ (in the unit of $k_{\text{B}}T$) in
canonical ensemble is defined by
\begin{eqnarray}
F&=&\frac{1}{z}\sum_{rr'}\left[ \chi_{\text{AB}}\phi_\text{A}(r)\phi_\text{B}(r')+\chi_{\text{AC}}\phi_\text{A}(r)\phi_\text{C}(r')+\chi_{\text{BC}}\phi_\text{B}(r)\phi_\text{C}(r')
\right] -\sum_r\left[ \omega_\text{A}(r)\phi_\text{A}(r)\right.\nonumber\\
&+&\left.\omega_\text{B}(r)\phi_\text{B}(r)+\omega_\text{C}(r)\phi_\text{C}(r)
\right] -\sum_r{\xi(r)[1-\phi_\text{A}(r)-\phi_\text{B}(r)-\phi_\text{C}(r)]}-n\ln{Q}.
\end{eqnarray}

Here, $\chi_\text{AB}$, $\chi_\text{AC}$, $\chi_\text{BC}$ are the Flory-Huggins
interaction parameters between different species. The $\phi_k(r)$ is
the volume fraction field of block specie $k$, which is independent
of the individual polymer configuration, and $\omega_k(r)$ is the
chemical potential field conjugated to $\phi_k(r)$. The $\xi(r)$ is
the potential field that ensures the incompressibility of the
system, also known as a Lagrange multiplier. $Q$ is the single chain partition function 
where,
\begin{equation}
Q=\frac{1}{N_\text{L}}\frac{1}{z}\sum_{r_N}\sum_{\alpha_N}G^{\alpha^{N}}(r,N|1).
\end{equation}
Where, $r_{N}$ and $\alpha_{N}$ denote the position and bond
orientation of the $s$-th segment of a copolymer,
respectively.  $z$ denotes lattice
coordination number.
The
end-segment distribution function $G^{\alpha _{s}}(r,s|1)$ presents
the statistical weight of all possible configurations staring from
segment 1, which can be located in any position within the lattice,
ending at segment $s$ at site $r$. Following the scheme of Scheutjens and Leemakers \cite{Leer1988}, $G^{\alpha _{s}}(r,s|1)$ satisfies the following
recurrence relation:
\begin{equation} G^{\alpha _{s}}(r,s|1)=G(r,s)\sum_{r_{s-1}^{\prime
}}\sum_{\alpha _{s-1}}\lambda _{r_{s}-r_{s-1}^{\prime }}^{\alpha
_{s}-\alpha _{s-1}}G^{\alpha _{s-1}}(r{'},s-1|1),
\label{free}
\end{equation}
 $G(r,s)$ is the weight factor of
the free segment. It is expressed as $G(r,s)=\exp[-\omega
_{\beta}(r_{_{s}})],\  s\in \beta(\beta=\text{A,\,B,\,C})$. Here, $r_{s-1}^{\prime
}$ and $\alpha_{s-1}$ denote the position and bond
orientation of the $s-1$ segment of a copolymer,
respectively. For all the values of $\alpha _{1}$, the initial condition is
$G^{\alpha _{1}}(r,1|1)=G(r,1)$.

The transfer matrix $\lambda$ depends only on the
chain model used, and this paper adopts an inflexion chain model. For a
coil subchain,
%
\begin{eqnarray}
\lambda _{r_{j,s}-r{'}_{j,s-1}}^{\alpha _{j,s}-\alpha
_{j,s-1}}=\left\lbrace 
\begin{array}{ll}
0, & \alpha _{j,s}=-\alpha_{j,s-1}\\
{1}/(z-1), & \text{otherwise}\,,
\end{array}\right.
\end{eqnarray} 
$r{'}$ denotes the nearest neighboring site of $r$.
$\alpha_{j,s}$ can choose all of the possible bond orientations,
which depends on the selected lattice model.
For a rigid subchain,
%
\begin{eqnarray}
\lambda _{r_{j,s}-r{'}_{j,s-1}}^{\alpha _{j,s}-\alpha _{j,s-1}}=\left\lbrace 
\begin{array}{ll}
1, & \alpha_{j,s}=\alpha _{j,s-1}\\
0, & \text{otherwise}.
\end{array}\right.
\end{eqnarray} 
Another end-segment
distribution function $G^{\alpha _{s}}(r,s|N)$ satisfies the
following recurrence relation:
\begin{equation}
G^{\alpha
_{s}}(r,s|N)=G(r,s)\sum_{r_{s+1}^{\prime }}\sum_{\alpha
_{s+1}}\lambda _{r_{s+1}^{'}-r_{s}}^{\alpha _{s+1}-\alpha
_{s}}G^{\alpha _{s+1}}(r',s+1|N).
\end{equation}
With the initial condition $G^{\alpha _{_N}}(r,N|N)=G(r,N)$ for all
the values of $\alpha _{_N}$.

Minimization of the free energy functional $F$ with respect to
$\phi_\text{A}$, $\phi_\text{B}$, $\phi_\text{C}$, $\omega_\text{A}$, $\omega_\text{B}$,
$\omega_\text{C}$ and $\xi(r)$ leads to the following SCFT equations:
\begin{equation}
\omega_A(r)=\frac{1}{z}\sum_{r'}\chi_\text{AB}\phi_\text{B}(r')+\frac{1}{z}\sum_{r'}\chi_\text{AC}\phi_\text{C}(r')+\xi(r),
\end{equation}
\begin{equation}
\omega_\text{B}(r)=\frac{1}{z}\sum_{r'}\chi_\text{BC}\phi_\text{C}(r')+\frac{1}{z}\sum_{r'}\chi_\text{AB}\phi_\text{A}(r')+\xi(r),
\end{equation}
\begin{equation}
\omega_\text{C}(r)=\frac{1}{z}\sum_{r'}\chi_\text{AC}\phi_\text{A}(r')+\frac{1}{z}\sum_{r'}\chi_\text{BC}\phi_\text{B}(r')+\xi(r),
\end{equation}
\begin{equation}
\phi_\text{A}{(r)}+\phi_\text{B}{(r)}+\phi_\text{C}{(r)}=1,
\end{equation}
\begin{equation}
\phi_\text{A}(r)=\frac{1}{N_\text{L}}\frac{1}{z}\frac{n}{Q}\sum_{s\in{\text{A}}}\sum_{\alpha_s}\frac{G^{\alpha_s}(r,s|1)G^{\alpha_s}(r,s|N)}{G(r,s)},
\end{equation}
\begin{equation}
\phi_\text{B}(r)=\frac{1}{N_\text{L}}\frac{1}{z}\frac{n}{Q}\sum_{s\in{\text{B}}}\sum_{\alpha_s}\frac{G^{\alpha_s}(r,s|1)G^{\alpha_s}(r,s|N)}{G(r,s)}\,,
\end{equation}
\begin{equation}
\phi_\text{C}(r)=\frac{1}{N_\text{L}}\frac{1}{z}\frac{n}{Q}\sum_{s\in\text{C}}\sum_{\alpha_s}\frac{G^{\alpha_s}(r,s|1)G^{\alpha_s}(r,s|N)}{G(r,s)}.
\end{equation}

In our calculations, the real-space method is implemented to solve
the SCFT equations in a cubic lattice with periodic boundary
conditions \cite{Chen2006,Xia2010}. The calculations begin from the
initial potential fields generated by random functions, and stop
when the free energy of the system changes within a tolerance of
$10^{-8}$. By comparing the system free energies from different random configurations, we obtained  stable phases in the system
of  coil/coil/rod ABC triblock copolymers.

\section{Result and discussion\label{sec3}}
 In our studies,  the effect of the asymmetric interactions between  different components on self-assembly is studied. In order to feature the influences of the chain rigidity and the entropy in the coil/coil/rod ABC triblock copolymers, we assume that the interaction between A and C components equals the interaction between B and C components,  and that the interaction between coil blocks is relatively small ($\chi_\text{AB} N=10$).  Furthermore, we only focus on the copolymer with the same relative length of A and B coil blocks, i.e., $f_\text{A} = f_\text{B}$. The degree of polymerization of the copolymers $N$ equals $20$. Thus, the ordered morphologies of the copolymers depend on two molecular parameters: $f_\text{rod}$ (the volume fraction of the rod block C) and $\chi_\text{AC} N$ (the interaction between A and C components). The calculations are
preformed in $N_\text{L} = 60^3 $ to $N_\text{L} = 80^3 $ lattices to make sure that the emergence of self-assembled structures is not
constrained by system size.
\begin{figure}[!t]
\centering
\includegraphics[width=7cm]{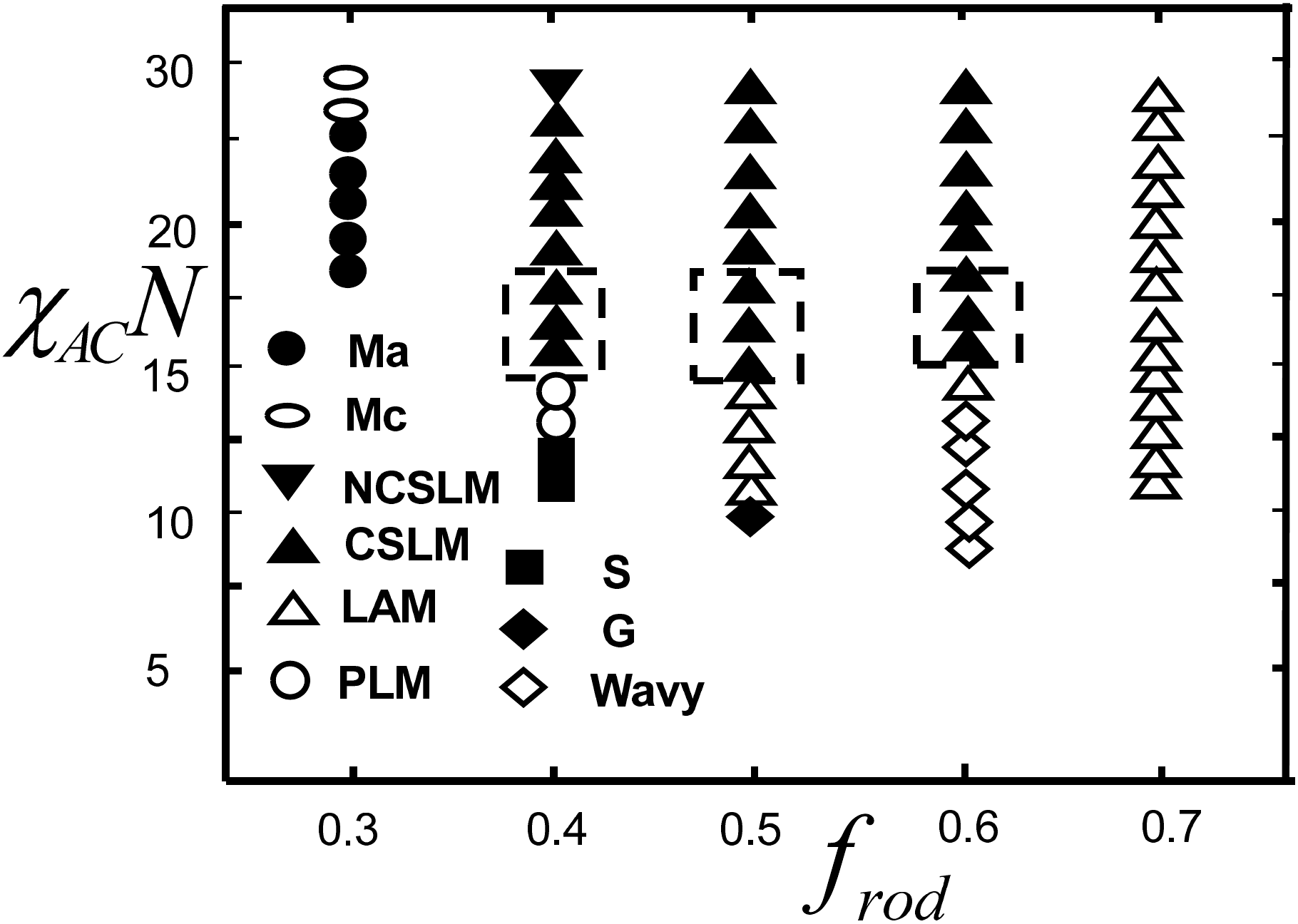} 
\caption{The phase diagram for linear  coil/coil/rod ABC triblock copolymers with $\chi_\text{AB} N=10$ ($\chi_\text{AC}N=\chi_\text{BC} N$): micelles A (Ma), micelles C (Mc), non-centrosymmetric  lamellae (NCSLM), centrosymmetric  lamellae (CSLM),  lamellae (LAM), perforated lamellae (PLM), strips (S), gyroid (G) and wavy.  The dashed lines in the phase diagram indicate the regions of the emergence of degenerate behavior of NCSLM and CSLM.
\label{phase}}
\end{figure}

\begin{figure}[!t]
\centering
\includegraphics[width=3.5cm]{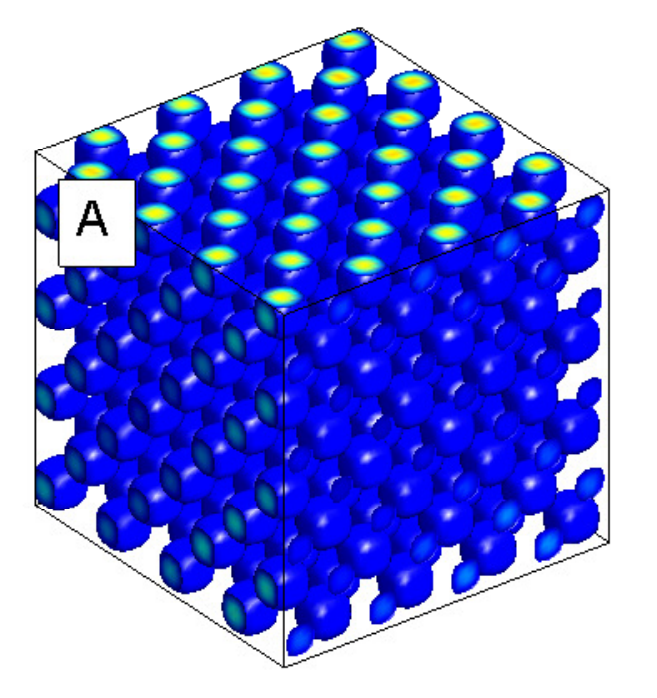}
\includegraphics[width=3.5cm]{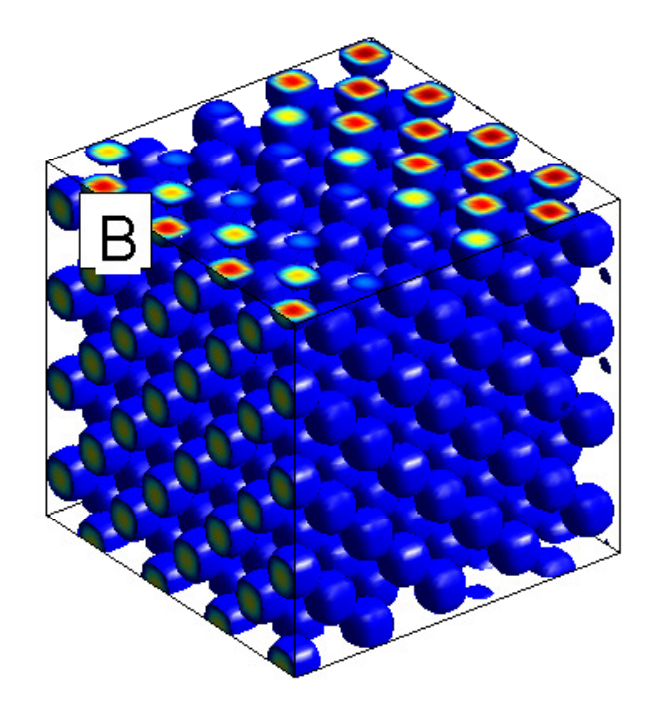}
\includegraphics[width=3.5cm]{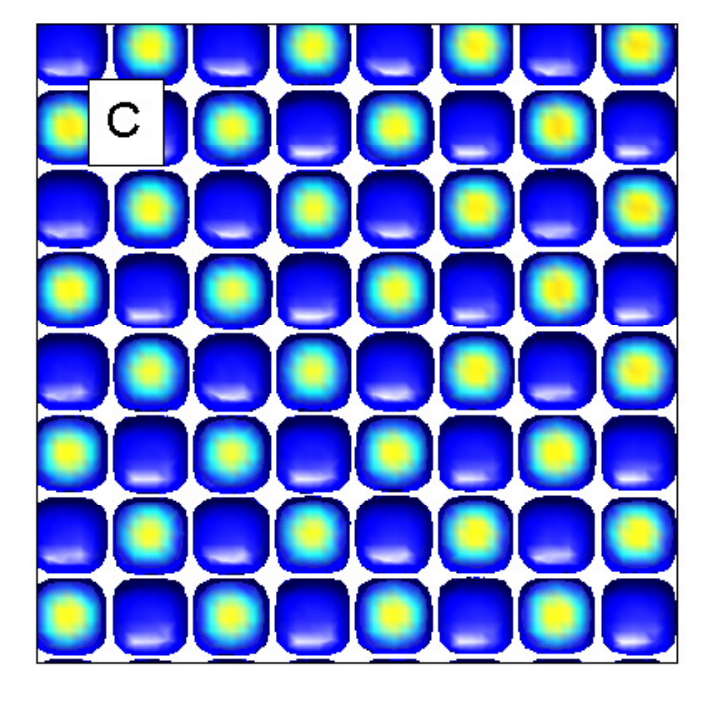}
\includegraphics[width=3.5cm]{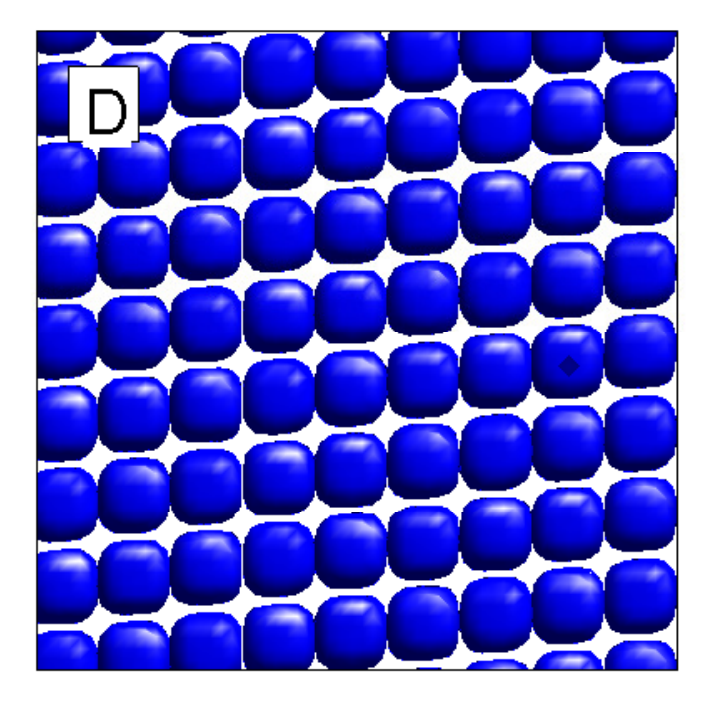}
\caption{(Colour online) The ordered structures observed at $f_\text{rod}=0.3$ (The vacant space is the components A and B which are eliminated for clarity. The following figures are alike). (A): micelles A, (B): micelles C, (C): The X-Z section of micelles A, (D): The X-Z section of micelles C.
\label{mice}}
\end{figure}
Figure~\ref{phase} shows the phase diagram constructed by $f_\text{rod}$ versus $\chi_\text{AC} N$. In addition to micelles, centrosymmetric lamellae (CSLM), lamellae, perforated lamellae, strips and gyroid, noncentrosymmetric lamellae (NCSLM) and wavy  lamellae are observed. Here, NCSLM and CSLM are degenerate structures. These phases are discussed below.

When $f_\text{rod}=0.3$, as shown in figure~\ref{mice}, two kinds of micelles are observed, i.e., micelle A and micelle C structures. These two kinds of micellar morphologies arrange in a different way within three mutually perpendicular sections of cubic lattice. In micelle A, micelles arrange along the coordinate axis within the three sections. The arrangement of micelles of micelle C within the two X-Y and Y-Z sections is similar to that of micelle A. In X-Z section, however, micelles with in micelle C does not arrange  along the coordinate axis. Micelle A is obtained as a stable structure at $\chi_\text{AC} N=18$. When $\chi_\text{AC} N$ is increased to $28$, micelle C is more stable than micelle A. For the same ratio of rod to coil blocks with the symmetric interaction and for the same relative length of coil blocks case, CSLM is only a stable phase. However, when the relative length of A blocks becomes small, micelle (A) structure is more stable than CSLM \cite{Xia2010}. It is demonstrated that the influence of the asymmetric interaction in the system with $f_\text{A}=f_\text{B}$ is to some extent similar to that of the asymmetry of the two coil blocks of the triblock copolymers with symmetric interactions, where the systems tend to make the coil blocks more free volume to maximize entropy. That is, the entropy  plays a more important role in coil/coil/rod ABC triblock copolymers  with $f_\text{A}=f_\text{B}$ when $\chi_\text{AB} N$ is relatively small.
\begin{figure}[!]
\centering
\includegraphics[width=3.5cm]{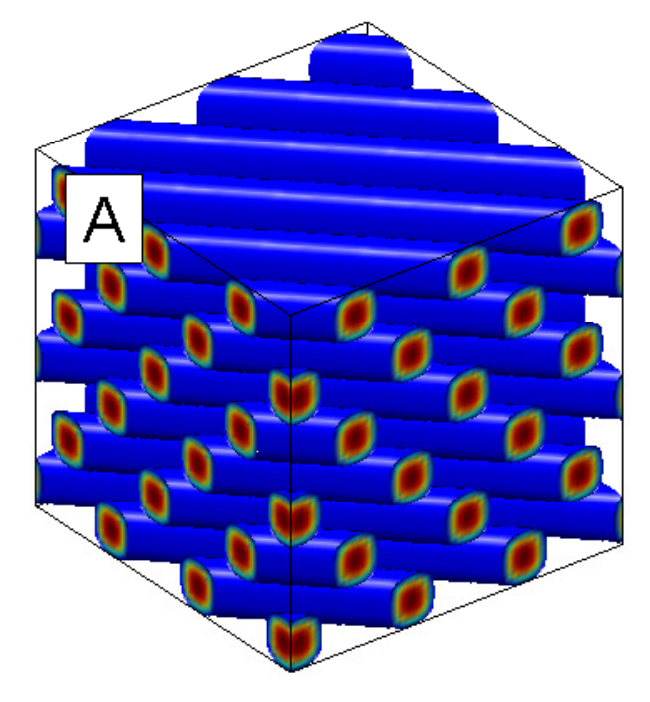}
\includegraphics[width=3.5cm]{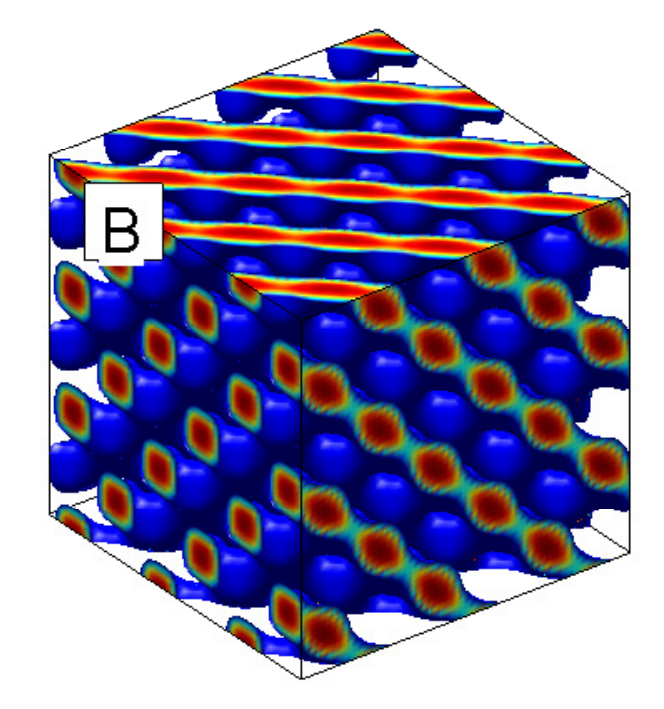}
\includegraphics[width=3.5cm]{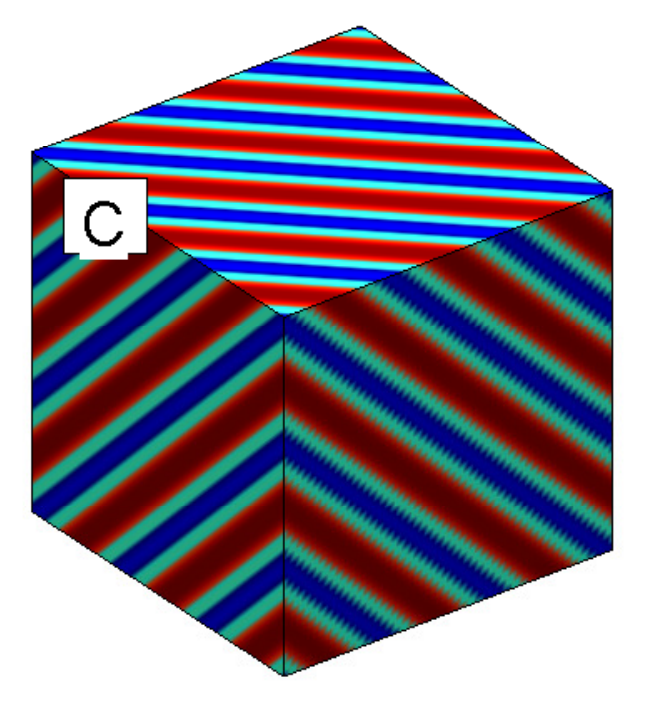}
\includegraphics[width=3.5cm]{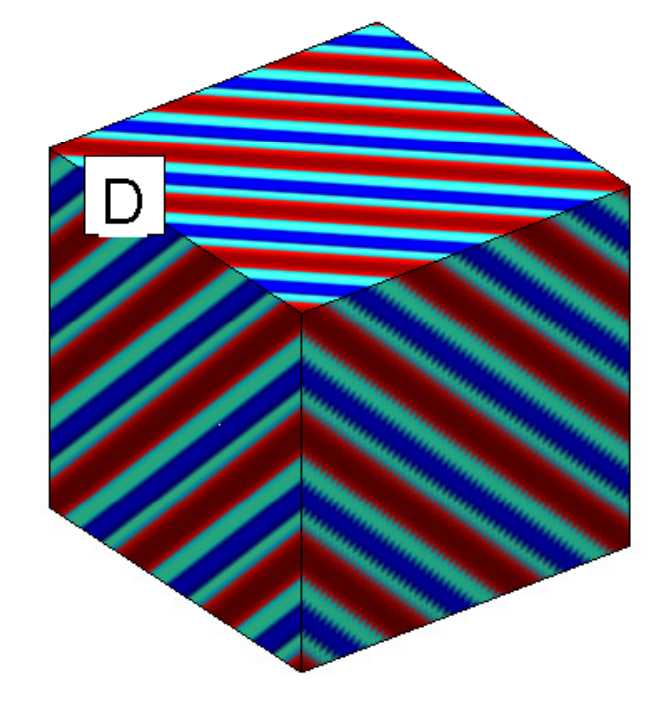}
\includegraphics[width=6.5cm]{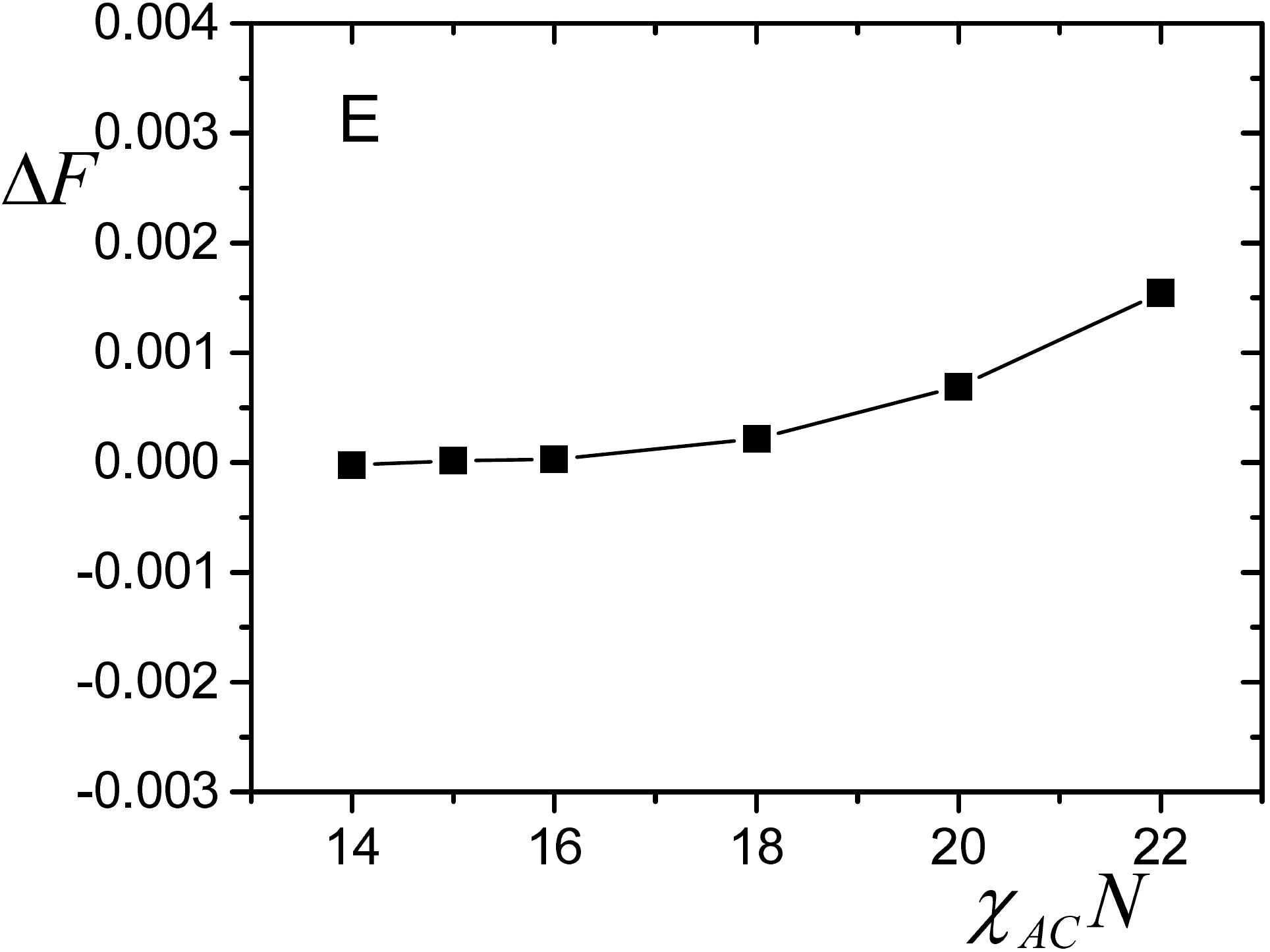}
\includegraphics[width=7cm]{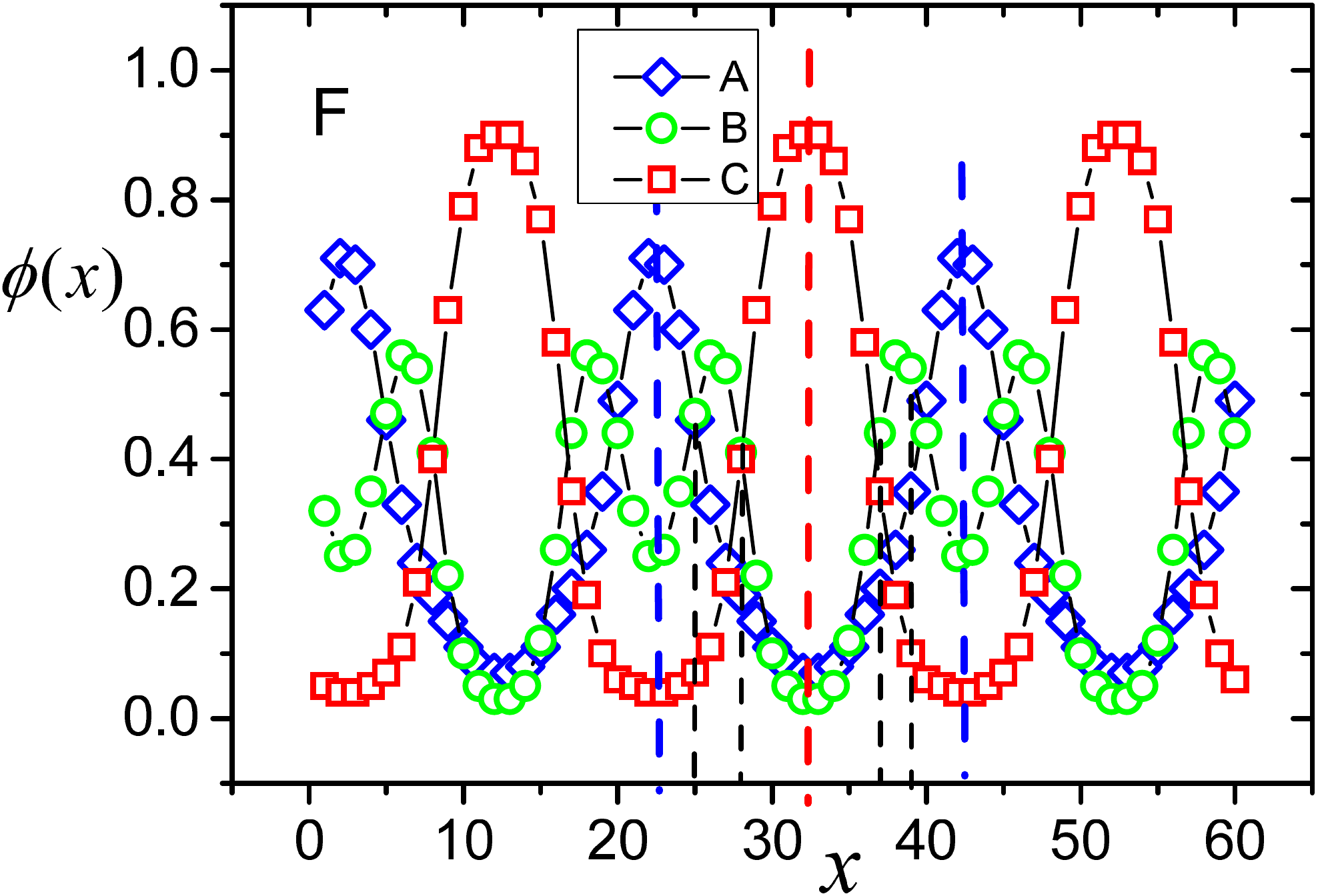}
\caption{(Colour online) The ordered morphologies observed at $f_\text{rod}=0.4$: (A): strips, (B): perforated lamellae, (C): centrosymmetric lamellae, (D): noncentrosymmetric lamellae, (E): the curve of the difference between the free energies of CSLM and NCSLM  with $\chi_\text{AC}N$, (F) : the distributions of volume fractions of three components along $a$ axis.\label{LM3}}
\end{figure}

At $f_\text{rod}=0.4$, as showed in figure~\ref{LM3}, strips and perforated lamellae  emerge as stable phases near order/disorder transition. When $\chi_\text{AC} N\geqslant 14$, NCSLM is observed. NCSLM is an ABCBA lamellar structure, though the width of the two B layers is different. NCSLM was observed in flexible ABCA and ABCD tetrablock copolymers\cite{Karim2004,Takano2003,Takano2003a} and in the blend of AC diblock and ABC triblock copolymers \cite{Wickham2001,Erukhimovich2010,Leibler1999,Amoskov1998,Tournilhac1992}. However, it is found for the first time in coil/rod block copolymers.  It is noted that the difference between the free energies of NCSLM and CSLM  is very small (no more than the order of magnitude of $10^{-5}$)  when $14\leqslant\chi_{AC }N<18$ [seen figure~\ref{LM3} (E)], i.e., CSLM and NCSLM are degenerate structures. Above the region of degenerate behavior in the phase diagram, the relative stability between NCSLM and CSLM changes with an increase in $\chi_\text{AC} N$. When $\chi_\text{AC} $ is increased, firstly CSLM has a lower free energy than NCSLM,  and then  NCSLM is more stable than CSLM. This is similar to the emergence of the stable micelle C at $f_\text{rod}=0.3$. These behaviors are in reasonable agreement with the cylindral/ spherical structure transition in coil/rod/coil ABA triblock with increasing the interaction between rod and coil blocks \cite{Chen12007}.  The mechanism  for the emergence of NCSLM will be discussed in detail  below.

\begin{figure}[!t]
\centering
\includegraphics[width=3.6cm]{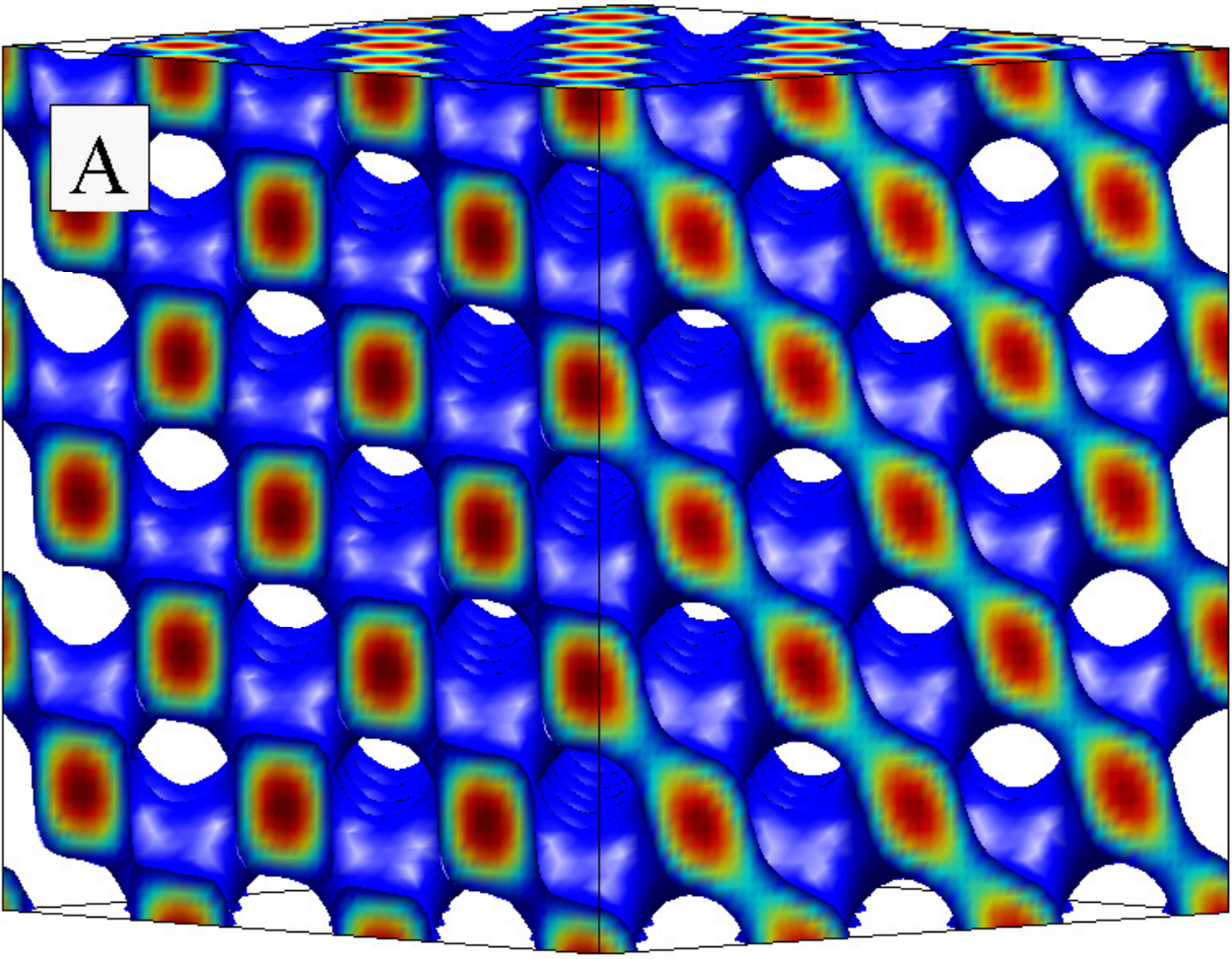}
\includegraphics[width=3.5cm]{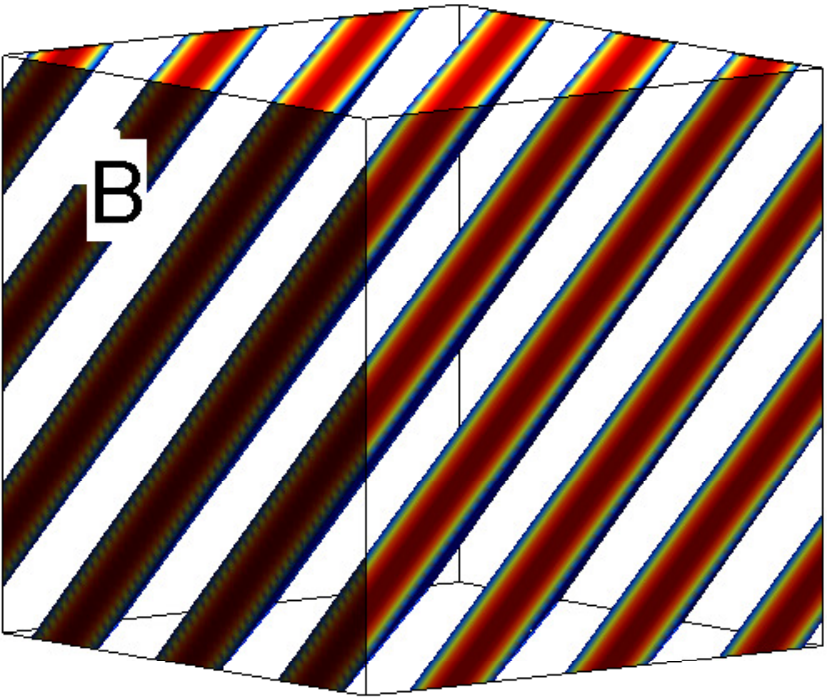}
\includegraphics[width=3.3cm]{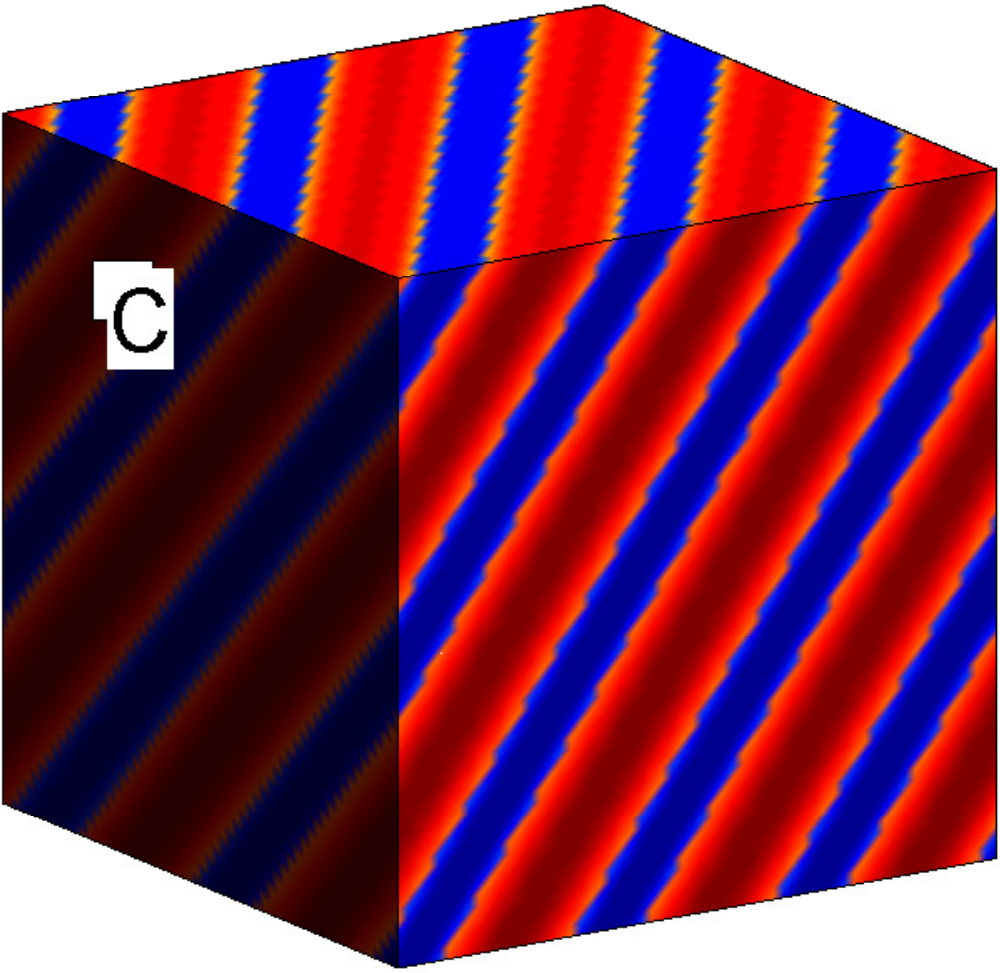}
\caption{(Colour online) The ordered structures observed $f_\text{rod}=0.5$ and $f_\text{rod}=0.6$ : (A): gyroid; (B): lamellae, (C):  wavy.
\label{Zg}}
\end{figure}

For $f_\text{rod}=0.5$, the gyroid/lamellae transition appears, which is consistent with coil/rod diblock copolymers \cite{Chen12006}. For large $\chi_\text{AC}N$, firstly NCSLM and CSLM are also observed as degenerate structures,  and then CSLM is only a stable phase. At $f_\text{rod}=0.6$, wavy is observed  [see figure~\ref{Zg}]. Wavy  is a morphology where a lamellae has  chevron-like defects within the interfaces between the coil and rod segments in the two sections of three mutually perpendicular sections.  Zigzag and wavy morphologies were regarded as the same structure in coil/rod diblock copolymers \cite{Horsch2010}. Much effort is made to determine and explain the stability of zigzag (wavy) morphology in the previous work  \cite{Chen12006, Li2001,Horsch2010,Gao2011}. When $8\leqslant\chi _{AC}N<14$,  wavy is more stable than lamellae in coil/coil/rod ABC triblock copolymers.  Compared with lamellae,  wavy, where it has  defects, is favorable in order to increase the space for junction points of coil/rod blocks to maximize the  entropy of the coils. This factor can be an important  reason for the stability of zigzaglike morphology when $\chi_\text{AC}N$ is small. Then, lamellae and CSLM become a stable phase sequently with the increase in $\chi _{AC} N$.

For intermediate rod fraction, the balance between internal energy and entropy is more delicate in the coil/rod block copolymers, which  is demonstrated by the rich phase behavior discussed above. When $\chi_\text{AB} N$ is relatively small, coil/coil/rod ABC triblock copolymers are similar to the coil/rod diblock copolymers. On the other hand, these two systems behave differently in a way. For the copolymer chains  assembled into layers, generally the stretchings in lamellae for coil/rod AC diblock and coil/coil/rod ABC triblock copolymers are different, since the stretching in triblocks is governed by the two kinds of  interaction between  A/B and B/C, rather than  only a/c interaction as in diblocks. Compared with coil/rod diblock copolymers, the stretchings from the A/B interface lead to a smaller free volume of coil block for the coil/rod triblock copolymers, and do not favor the entropy. Accordingly,  the emergence of more chevron-like defects is necessary to maximize the entropy through decreasing the  grafting density of the interfaces of rod and coil segments. Consequently, the stable wavy morphologies are observed at $f_\text{rod}=0.6$.  Seen from figure~\ref{LM3} (F), the emergence of NCSLM results from the asymmetric interfaces. A different degree of chain stretching to maximize the entropy leads to the emergence of an asymmetric interface. Naturally the emergence of NCSLM is deduced by the different chain stretching of coil blocks on the two sides of the domain formed rod blocks, which is similar to the case of  NCSLM in the blends of the flexible AC diblock and ABC triblock copolymers \cite{Tournilhac1992}. It is expected that the different degree of chain stretching is dependent on $\chi_\text{AC} N$. When $\chi_\text{AC} N$ (<18) is not so big, the effect of entropy from the less-stretched  coil blocks offsets that of the internal energy from greater contact between different components. Thus,  CSLM and NCSLM are degenerate structures. With an increase in $\chi_\text{AC} N$,  the influence of entropy weakens, and the internal energy markedly dominates the stability of assembled structures.  With  $\chi_\text{BC} N$ ($=\chi_\text{AC} N$)  increasing, the width of asymmetric interface  of  NCSLM decreases. The asymmetric interface becomes favorable to the internal energy. Therefore, NCSLM is more stable than CSLM. 

Experimental studies on  coil/coil/rod ABC triblock copolymers have  demonstrated a significant influence of the rigid block on self-assembly of block  copolymers. The introduction of rod-like polymers adds an additional geometrical asymmetry and the rod/rod interaction. In  ABC $\pi$-conjugated coil/coil/rod triblock copolymers, the  triple-lamellar/double-lamellar phase  transition emerges near the order/disorder transition. At the same time,  the ordered lamella/disorder transition and the smectic/isotropic transition occur at the same temperature, indicating that the rod/rod interaction between PPV rods plays a critical role in forming and stabilizing these lamellar structures \cite{Chang2011}. Although  SCFT lattice model does not sufficiently take account of rod/rod anisotropic interaction, it had predicted triple-lamellar/double-lamellar phase  transition \cite{Xia2010}. In order to maximize the influence of the incompatibility between coil and rod blocks, as well as the  block stiffness, on microphases separation, it is assumed that the coil/coil interaction, i.e.,  $\chi_\text{AB} N$, is relatively suitable  (corespongding to the order-disorder transition for coil/coil diblock copolymers). In other words, when  separation of microphases occurs, the effect of A/B interface on the B/C interface is weak, and it is favourable to the emergence of asymmetic interface. When $\chi_\text{AB} N$  is increased to 12 or decreased to 8 at $f_\text{rod}=0.4$, the region of the phase diagram of  stability  and degenerate behavior apparently decreased.  The emergence of  the NCSLM  is very much dependent on  $\chi_\text{AB} N$.  In  $\pi$-conjugated coil/coil/rod ABC triblock copolymers containing a mesogen-jacketed liquid crystalline polymer, the core-shell hexagonally packed cylinder/lamella transition emerges with increasing the temperature, \cite{Pan2016}  and the transition sequence  is in reasonable agreement with that of  NCSLM/CSLM transition. Consequently, it  is  expected that NCSLM may be observed in the experiment.

 Although the excluded volume effects are neglected, SCFT have been successful at describing the structural  properties of simple and complex polymers \cite{Muller2013}.
In the coil/coil AB diblock copolymers, the $\chi_\text{AB} N$ vs $f$ (the volume fraction of A block ) phase diagram calculated by SCFT \cite{Matsen1994} is well confirmed by the experimental study \cite{Hillmyer1997}. By the simulations of different continuum bead-spring models, it was found that although SCFT provides inaccurate predictions for order-disorder transition because of the neglect of fluctuation \cite{Duchs2003}, it gives surprisingly accurate predictions for the free energy of the ordered lamellar phase, suggesting that SCFT may provide  more reliable predictions for order/order transitions \cite{Glaser2014}. 
Furthermore, Chremos et al. \cite{Chremos2014} developed a coarse-graining methodology for mapping specific block copolymer systems to bead-based models. 
It correctly reproduced the characteristic domain morphologies of poly(styrene-b-methyl methacrylate) over a broad range of $ f$ and $\chi_\text{AB} N$ values, agreeing well with the results from SCFT calculations. They had obtained semi-empirical temperature-related equations that link $\chi$ to the model parameters, and predicted that the order-disoder temperature of   poly(styrene-b-methyl methacrylate)  is also in good agreement with the reported experimental results \cite{Chremos2014}.

\section{Conclusion and summary\label{sec4}}
In this work,  the self-assembly of coil/coil/rod ABC triblock copolymers is studied using a self-consistent field of lattice model under asymmetric interaction between different coil/rod components, given a small and fixed interaction between coil components. In addition to micelles, centrosymmetric lamellae, lamellae, perforated lamellae, strips and gyroid, noncentrosymmetric lamellar and wavy morphologies were observed as stable structures. For intermediate rod fraction, the degenerate behavior for NCSLM and CSLM was observed.  It is found that the entropy of chain conformation plays an important role in the emergence of NCSLM.  It is expected that the stability of NCSLM is related to the chain architecture. This will be studied in the further work.

\section*{Acknowledgements} 
This research is financially supported
by the National Nature Science Foundations of China (21564011) and
the Inner Mongolia municipality (2017MS(LH)0211).

\ukrainianpart
\title
{Самоскупчення в блокових кополімерах стрижень/клубок: вироджена поведінка  в умовах необмеженості %
}
\author{Кс.-Г. Ган, Н. Ліянг, Г. Шанг }
\address{Школа природничих наук, Унiверситет науки і технологій Внутрiшньої Монголiї, Баоту 014010, Китай }

\makeukrtitle 

\begin{abstract}
\tolerance=3000%
Самоскупчення  блокових кополімерів, що складаються з жорстких блоків, привертає значну увагу завдяки його багатій фазовій поведінці та потенційним можливостям використання у різноманітних пристроях. У даній роботі, за умов асиметричних взаємодій між компонентами клубок/стрижень, досліджується самоскупчення 
ABC триблокових кополімерів	 клубок/клубок/стрижень з використанням самоузгодженого поля граткової моделі.
На додаток до міцел, центросиметричних ламел,  ламел, перфорованих ламел, смуг та гіроїдів, спостерігаються нецентросиметричні ламели  та хвилясті морфології в якості стабільних фаз. Фазова діаграма взаємодії між стрижневими і клубковими компонентами  відносно фракції стрижнів побудована за умови фіксованої взаємодії між клубковими компонентами. Для проміжної стрижневої фракції спостерігається вироджена поведінка. Центросиметричні ламели і  нецентросиметричні ламели  --- вироджені структури. Встановлено, що ентропія ланцюгової конформації відіграє важливу роль у цій багатій поведінці. Запропоновано механізм дії виродженої поведінки в  блокових кополімерах стрижень/клубок в умовах необмеженості. Дане дослідження представляє дещо новий погляд на вироджену поведінку   блокових кополімерів, а це, у свою чергу, може запропонувати теоретичні підвалини для споріднених експериментів.

\keywords  блоковий кополімер стрижень/клубок, вироджена поведінка, самоузгоджене поле

\end{abstract}

  \lastpage
 \end{document}